\documentclass[aps,prm,reprint]{revtex4-2}

\usepackage{mathtools}
\usepackage{bm}
\usepackage{hyperref}
\hypersetup{colorlinks=true, citecolor=blue, urlcolor=blue, linkcolor=blue}

\begin{document}
\title{Finite-temperature grain boundary phases, transformations, and diagrams in alloys}

\author{Flynn Walsh}
\email{walsh40@llnl.gov}
\author{Timofey Frolov}
\email{frolov2@llnl.gov}
\affiliation{Lawrence Livermore National Laboratory, Livermore, California 94550, USA}

\begin{abstract}

Current understanding of alloy grain boundaries is limited by the incompleteness of atomistic simulation techniques, which either neglect entropy or artificially constrain the atomic density.
This study introduces Monte Carlo simulations that simultaneously sample all the microscopic degrees of freedom in complex interfaces: position, configuration, and atom number.
The method enables rigorous predictions of finite-temperature grain boundary structures, phase transformations, and diagrams in multicomponent materials.
Simulations of W-V alloys demonstrate the necessity of the approach by uncovering a range of previously inaccessible phase transitions including condensation of interstitial solutes and transformation of dislocation structure.
\end{abstract}
\maketitle

\section{Introduction}

Grain boundaries (GBs) control material properties ranging from fracture resistance to magnetic coercivity \cite{watanabe11,liu21b,quirk24}.
They are also scientifically fascinating as interface structures can be understood as thermodynamic phases that transform with variables such as temperature or composition \cite{frolov15}.
While advances in electron microscopy have enabled the atomic scale characterization of increasingly complex alloy GBs \cite{yu17,zhu17a,zhao19a,ahmadian21,futazuka22,seki23,zhou25,devulapalli24}, predicting the precise structures of multicomponent interfaces remains a major theoretical challenge.

Conventional alloy simulations combine isothermal-isobaric molecular dynamics with Monte Carlo (MC) swaps to sample configurational degrees of freedom for a fixed number of atoms, $N$.
This approach can capture phenomena such as substitutional segregation and certain chemical orderings, but fixing $N$ imposes an arbitrary interface density that predetermines the possible atomic structures.
The importance of varying $N$ has been extensively explored at 0\,K in both pure elements \cite{phillpot92,zhu18,chen24} and alloys \cite{chua10,banadaki18,devulapalli24,masuda25}, but accounting for temperature has traditionally required large-scale open-surface simulations \cite{frolov15a} that are ultimately more qualitative than quantitative.

More recently, Ref. \cite{walsh26} proposed using an extended system of real and fictitious particles to identify plausible sites for gradual insertions and deletions under some chemical potential, $\mu$.
When combined with volume sampling at constant pressure ($P$) and temperature ($T$), this approach enabled the simulation of simple elemental defects in a $\mu PT$ or ``unconstrained'' ensemble  \cite{campa18,latella21}.
However, the thermodynamic coupling between chemistry and interface density---which we show to be critical---has been little explored due to the lack of available methods.

This study combines unconstrained MC, new configurational sampling techniques, and chemical potential calculations to create a general framework for simulating multicomponent GBs.
W-V, of interest for fusion applications, is investigated as a simple example system that forms a completely miscible bcc solution.
Even then, unconstrained simulations reveal a wide range of interface phase transitions involving sharp discontinuities, gradual filling, chemical (dis)ordering, and complete structural rearrangements including changes in dislocation structure.
The frequency and diversity of alloying phase transformations demonstrate the necessity of the new approach while highlighting the complexity of interface phenomena beyond substitutional segregation.

\section{Results}

\begin{figure}
  \includegraphics[width=\linewidth]{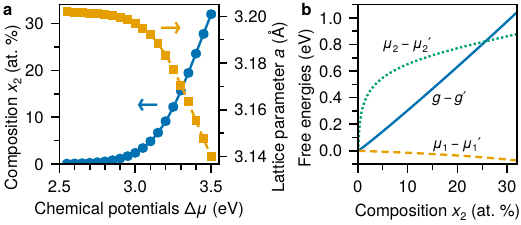}
\caption{
  \label{fig:sgcmc}
  Bulk properties are calculated as inputs for interface simulations; an example is shown for bcc W-V at 1600\,K.
  (a) The solute concentration, $x_2$, and lattice parameter, $a$, are equilibrated across a range of chemical potential differences, $\Delta\mu = \mu_2 - \mu_1$. 
  (b) Integrating $\Delta\mu(x_2)$ from the dilute limit provides the total free energy, $g$, and individual chemical potentials (see text).
}
\end{figure}

\subsection{Determining chemical potentials}

Unconstrained simulations first require the calculation of absolute chemical potentials that characterize the abstracted bulk phase(s).
For a given fractional composition, $x_2 = 1 - x_1$, chemical potentials can be related to the Gibbs free energy per atom as $g = G/N = \mu_1 x_1 + \mu_2 x_2$, assuming negligible point defects.
As an absolute reference, the free energy of an elemental solid, $\mu_1^e = g_1^e$, is first calculated by integrating from an Einstein crystal \cite{freitas16}.
Alloy free energies are then determined in an ensemble that fixes $\Delta \mu = \mu_2 - \mu_1$ in addition to $N$, $P$, and $T$ such that $dg = \Delta \mu \, dx_2$.
While the elemental composition of $x_1=1$ corresponds to the limit of $\Delta \mu \rightarrow - \infty$, we assume $\mu_1' = \mu_1^e$ under very dilute $x_2' \ll x_1'$, which is supported by later calculations.
Then, for $\Delta \mu'$ producing $x_2'$,
\begin{equation}
  \label{eq:gx2p}
  g(x'_2) = \mu_1' x_1' + \mu_2' x_2' = \mu_1^e + \Delta\mu' x_2'.
\end{equation}
Additional chemical potentials are calculated by integrating $g(x_2) = g(x_2') + \int_{x_2'}^{x_2} dg$ and applying Eq. (\ref{eq:gx2p}) as
\begin{equation}
  \label{eq:mu}
  \begin{split}
  \mu_1(x_2) &= g(x_2) - \Delta\mu(x_2) \cdot x_2, \\
  \mu_2(x_2) &= g(x_2) + \Delta\mu(x_2) \cdot (1-x_2).
  \end{split}
\end{equation}
A similar, if slightly more involved, procedure is possible with additional components.

Figure \ref{fig:sgcmc} demonstrates the calculation of absolute chemical potentials from $\Delta\mu(x_2)$ in bulk W-V alloys.
MC simulations equilibrated $x_2$ (circles) by trialing displacement trajectories, volume dilations/contractions, and direct chemistry changes in a bcc lattice.
The bulk lattice parameter (squares), which sets GB area, is also recorded.
Both quantities are fit to polynomials.
Figure \ref{fig:sgcmc}(b) plots integrated free energy and chemical potentials from Eq. (\ref{eq:mu}), relative to the dilute reference at $x_2' = 0.001027$ where $\Delta\mu' = 2.55$\,eV.

With $\mu(x)$ tabulated from $x_2'$ to $x_2$, $\mu P T$ interface simulations can explore how alloying transforms GB structures.
As described in Methods, MC simulations trialed displacements, insertions, deletions, transmutations, and grain shifts in one continuous Markov chain.
To minimize hysteresis effects, independent MC simulations were initialized far from equilibrium on grids of $x_2$ and $T$.
Phase transitions appeared sharply and consistently, validating the approach while providing input for GB phase diagrams.

\subsection{Example interface phase transformation}

\begin{figure}
  \includegraphics[width=\linewidth]{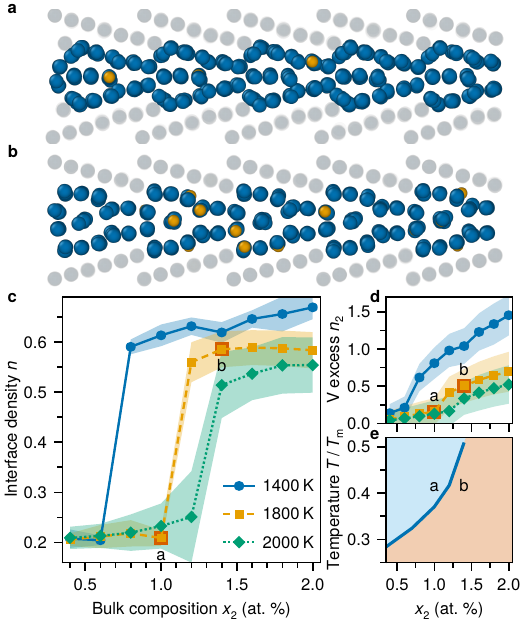}
\caption{
  \label{fig:gb552}
  A $\Sigma27(552)[1\bar{1}0]$ W-V GB demonstrates a clear discontinuous phase transformation upon dilute alloying.
  (a) Pure W forms an $n=0.2$ structure, which can only host dilute V.
  (b) Additional alloying induces a transformation to the $n=0.6$ structure.
  (c) The transformation is reflected in the interface density, which abruptly changes with bulk composition. Bands indicate ensemble fluctuations.
  (d) Segregation measured as the planar fraction of excess V.
  (e) Interface phase diagram.
}
\end{figure}

The $\Sigma$27 GB considered in Ref. \cite{frolov18} provides a clear demonstration.
Figure \ref{fig:gb552}(a) depicts the structure predicted for elemental W, which can host dilute V substitutions.
However, additional alloying transforms the GB into the structure shown in Fig. \ref{fig:gb552}(b).
The transformation is characterized by an abrupt change in interface density, which is plotted in Fig. \ref{fig:gb552}(c) for several temperatures.
The density is defined as the fraction of a plane parallel to the interface---for a symmetric GB cell with one parallel plane containing $N_\parallel$ atoms,
\begin{equation}
  n = (N \bmod  N_{\parallel} )/ N_{\parallel}.
\end{equation}
The excess V content, which measures segregation, can be similarly quantified as
\begin{equation}
  n_2 = (N_2 - x_2 N) / N_{\parallel},
\end{equation}
which is plotted in Fig. \ref{fig:gb552}(d).
While $n$ is periodic in $[0,1)$, $n_2$ has no specific limit.
The new phase, with $n\approx0.6$, accommodates several times more V than the elemental $n=0.2$ phase, pointing toward the transition's driving force.
Higher temperatures increase bulk solubility and reduce GB segregation, which shifts the transformation to greater V concentrations.
Figure \ref{fig:gb552}(e) plots the grain boundary phase diagram assessed from density discontinuities.

\subsection{Multiple interface phase transformations}

\begin{figure}
  \includegraphics[width=\linewidth]{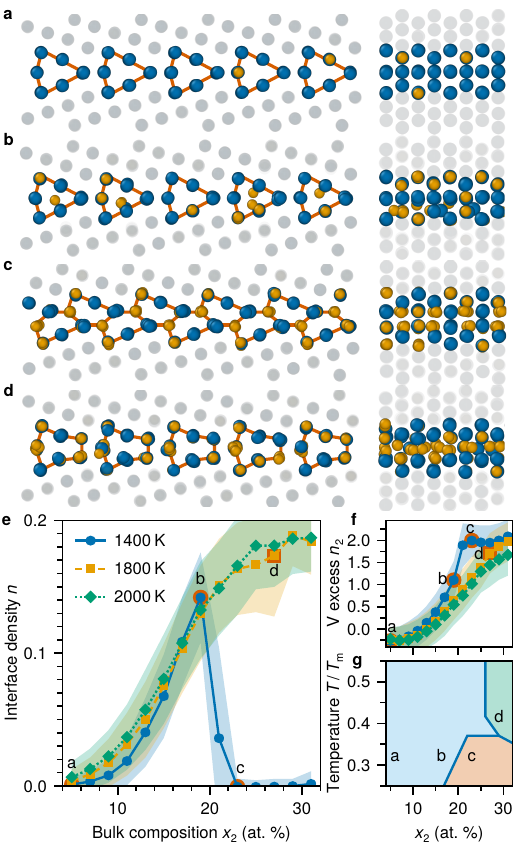}
\caption{
  \label{fig:gb210}
  A $\Sigma5(210)[001]$ W-V GB hosts three competing interface phases, plus interstitials.
  (a) The classic kite structure, which is stable under dilute alloying. Panels on the right show partial transverse perspectives.
  (b) Structurally disordered interstitials in the kite structure.
  (c) Asymmetric chemically ordered phase.
  (d) Chemically disordered split-kite phase.
  (e) The interface density vs. bulk composition, showing both phase transitions.
  (f) Segregation measured as the planar fraction of excess V.
  (g) Interface phase diagram.
}
\end{figure}

Unconstrained MC also reveals how certain boundaries can undergo multiple phase transformations at different temperatures and compositions.
Figure \ref{fig:gb210}(a) depicts the classic kite structure predicted for the $\Sigma 5(210)[001]$ symmetric tilt boundary in W.
Interestingly, unlike the previous example, segregation is initially negative at all temperatures, indicating that boundary substitution is unfavorable.
Positive segregation only occurs when solute atoms begin occupying interstitial sites at about 7\% V.
Figure \ref{fig:gb210}(b) illustrates how transient V occupies a variety of configurations within the interiors of kite columns.

Further alloying induces one of two phase transformations depending on temperature.
Below about $1600$\,K, ${\sim}18$\% bulk V causes the boundary to adopt the asymmetric ordered structure depicted in Fig. \ref{fig:gb210}(c).
This new phase replaces interstitial solutes with chemical decoration, causing the interface density to drop to $n=0$ in Fig. \ref{fig:gb210}(e).
Because the structure is chemically ordered, Fig. \ref{fig:gb210}(f) shows $n_2 \approx 2$ across a range of bulk compositions.
The ordering does not occur above ${\sim}1600$\,K, but kites continue to fill with interstitials up to ${\sim}26$\% bulk V, at which point they transform into the split-kite structure depicted in Fig. \ref{fig:gb210}(d).
Even though the transition does not dramatically change density or segregation, the physical transformation appears discontinuous.
Figure \ref{fig:gb210}(g) diagrams the three phases together.

\subsection{Interstitial-driven phase transformations}

\begin{figure}
  \includegraphics[width=\linewidth]{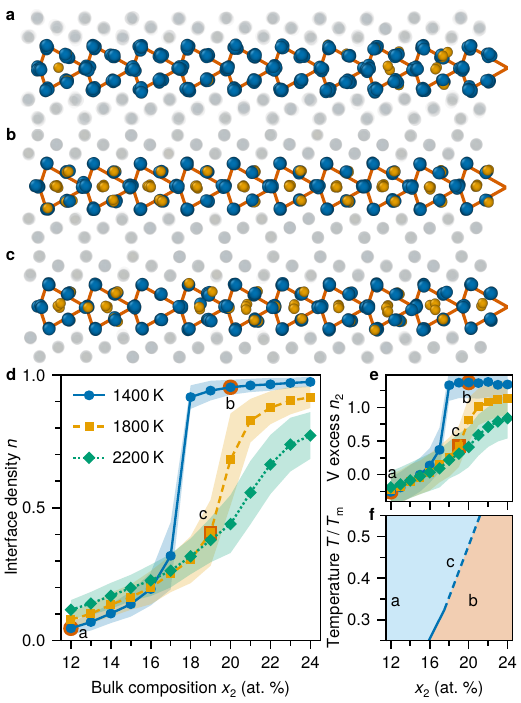}
\caption{
  \label{fig:gb310}
  Interstitial solutes appear to undergo both continuous and discontinuous phase transformations in a $\Sigma5(310)[001]$ W-V GB.
  (a) The kite structure of pure W, with some scattered V interstitials.
  (b) Discontinuous ordering of interstitials as an additional atomic layer.
  (c) More continuous condensation of interstitials at higher temperatures.
  (d) Interface density across the phase transition, highlighting the two regimes.
  (e) Segregation measured as the planar fraction of excess V.
  (f) Interface phase diagram, where the dashed line indicates the more continuous transition.
}
\end{figure}

In the $\Sigma5(210)[001]$ GB, phase transformations occur before interstitials completely fill kites.
In other boundaries, such as the related $\Sigma5(310)[001]$ structure, unconstrained MC predicts that interstitial solutes can undergo phase transformations independently from the host structure.
Below about 18\% V, solute atoms occupy kite centers in a similar transient manner, as depicted in Fig. \ref{fig:gb310}(a).
However, rather than transforming the entire boundary, fluid-like solutes appear to condense in a manner that depends on temperature.
At 1400\,K, V interstitials abruptly form the complete structure shown in Fig. \ref{fig:gb310}(b).
In contrast, the high-temperature transition appears continuous within the resolution of the simulations, passing through structures such as the partial filling depicted in Fig. \ref{fig:gb310}(c).
The two distinct types of transformation, which suggest a GB critical point, are reflected in the interface density in Fig. \ref{fig:gb310}(d) and the normalized segregation of Fig. \ref{fig:gb310}(e).
The phase diagram in Fig. \ref{fig:gb310}(f) indicates discontinuities in $n$ with solid lines and estimates the continuous crossover with a dashed line.

\subsection{Low-angle phase transformations and dislocations}

\begin{figure}
  \includegraphics[width=\linewidth]{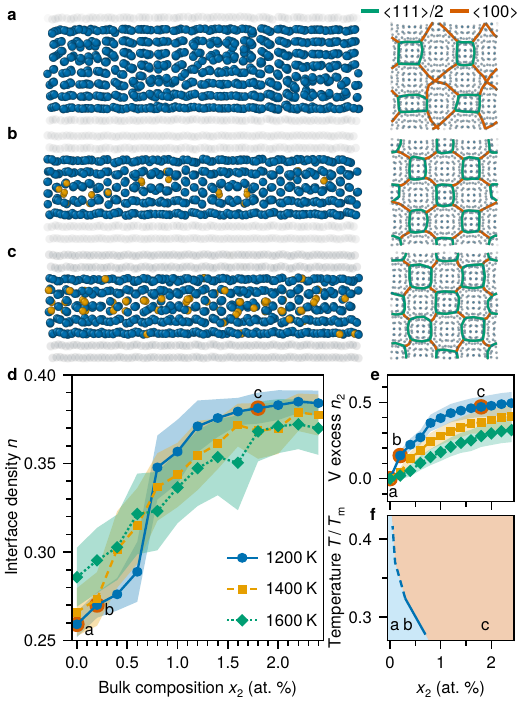}
\caption{
  \label{fig:gb001}
  Alloying phase transformations alter the dislocation content of a $\Sigma41(001)[001]$ low-angle twist GB in W-V.
  (a) The wavy boundary structure of pure W.
  (b) Dilute V segregates to the boundary, flattening the structure.
  (c) Additional V completely flattens the boundary in an apparent phase transformation.
  (d) Interface density as a function of concentration at different temperatures.
  (e) Segregation measured as the planar fraction of excess V.
  (f) Interface phase diagram.
}
\end{figure}

The previous three examples have focused on symmetric tilt boundaries, which provide familiar references that can appear in other structures due to faceting.
Unconstrained MC also reveals alloying phenomena in very different GBs such as a $\Sigma41(001)[001]$ low-angle twist boundary, which was motivated by a previous elemental investigation \cite{NOTE1}.
In pure W, the potential predicts the complex three-dimensional structure presented in Fig. \ref{fig:gb001}(a).
After dethermalization, this boundary can be interpreted as a square grid of dislocations with $\frac{1}{2}\langle111\rangle$ Burgers vectors that are diagonally connected by intersecting segments with $\langle100\rangle$ Burgers vectors.

In contrast to the tilt GBs, dilute V immediately segregates to the boundary core, as depicted in Fig. \ref{fig:gb001}(b) for a 0.2\% V alloy at 1200\,K.
Segregation induces a significantly flatter structure that maps to a distinct dislocation topology with an additional grid of $\frac{1}{2}\langle111\rangle$ loops appearing at the intersections of $\langle100\rangle$ segments.
At 1200\,K, a bulk composition of 0.7\% V completely flattens the boundary to the structure shown in Fig. \ref{fig:gb001}(c). 
This transformation is associated with a grain translation and density discontinuity that is visible in Fig. \ref{fig:gb001}(d).
A less pronounced discontinuity occurs around 0.3\% V at 1400\,K.
No clear jump is visible at 1600\,K, where the pure W boundary displays slightly higher equilibrium density.
At all temperatures, density and segregation plateau around 2\% bulk V.

\section{Discussion}
These four examples not only demonstrate the generality of unconstrained MC, but also provide a view of atomic structure that is somewhat richer than conventional perspectives.
While previous work has emphasized how chemistry can vary among boundaries \cite{obrien18}, the present results show how the underlying nature of segregation is not only variable but often structurally transformative.
This point contrasts with popular modeling approaches that calculate one or more segregation energies for individual solute atoms in substitutional or interstitial sites.

Previous studies have predicted some of the same structures as unconstrained MC, but conventional sampling techniques systematically underestimate the frequency and diversity of alloying phenomena due to density constraints. The widely studied $\Sigma5(210)[001]$ boundary considered in Fig. \ref{fig:gb210} provides an instructive example.
Using semi-grand canonical MC, Ref. \cite{yang18} predicted the same asymmetric ordered phase in Mo-Ni, which is possible because the two structures happen to share the $n=0$ interface density.
The persistence of both phases in open simulations validates the intuition underlying this approximation, but a closed ensemble could not predict the transient kite filling in Fig. \ref{fig:gb210}(b) or the split-kite transformation in Fig. \ref{fig:gb210}(d).

The $\Sigma5(310)[001]$ GB in Fig. \ref{fig:gb310}, which also has two $n=0$ structures, tells a similar story.
The filled structure in Fig. \ref{fig:gb310}(b) resembles conventional segregation patterns with integer solute layers \cite{cantwell14}, but the condensation is considerably less well explored despite similarities to bulk interstitial phenomena \cite{khachaturyan08}.
The practical implications of transient interstitials are unclear---they would be challenging to resolve microscopically and may also form two-phase regions in real boundaries---but the effect on GB diffusion kinetics should be significant.

Neither of the $\Sigma27(552)[1\bar{1}0]$ GB phases in Fig. \ref{fig:gb552} could be predicted by a naive semi-grand canonical ensemble, though the two structures have been previously identified as distinct energetic minima in pure W \cite{frolov18}.
Free energy calculations \cite{freitas18,choi25} could be adapted to compare their finite-temperature stabilities, but this approach becomes significantly more involved in multicomponent systems and neglects subtle changes in density.
The $\Sigma41(001)[001]$ boundary considered in Fig. \ref{fig:gb001} seems considerably less tractable as the transformed phases appear not as distinct minima, but continuously evolve with segregation in a manner that would require extensive optimization at each composition.

In all cases, unconstrained MC provides a far more direct and robust approach for interface structure prediction.
Nonetheless, the method would benefit from some refinement, particularly with regard to efficiency.
Collecting statistics for quantitative analysis requires $10^7$--$10^8$ MD steps per temperature and composition, which is more than is ideal for systematic studies, if very feasible with semi-empirical potentials.
The basic problem is that insertions and deletions are much more expensive than conventional MC moves and extensive in boundary area.
Since efficiency roughly follows homologous temperature, and V induces ordering at temperatures far below the W melting point, the present example of W-V may be particularly challenging.

\section{Conclusion}

This study has introduced unconstrained MC for multicomponent $\mu PT$ ensembles, providing a general method for calculating GB phase diagrams that are complete in position, configuration, and number degrees of freedom.
The approach has been demonstrated on four W-V GBs, revealing a wide array of interface phenomena that could not be rigorously modeled with existing techniques.
Predicted phase transitions include abrupt jumps in density, discontinuous transformations at constant density, the condensation of interstitials, interfacial chemical ordering, and changes in dislocation structure.
The varied nature of these effects poses a significant challenge for engineering GB structures, but the method provides an important tool for understanding them.

\section{Methods}

\subsection{Monte Carlo}
A multicomponent $\mu PT$ ensemble consists of $N$-particle microstates with positions $\bm{x}^N$ and types $\bm{\sigma}^N$. 
Potential energy can be expressed as $U(\bm{\xi}^N)$ for $\bm{\xi}^N = (\bm{x}^N,\bm{\sigma}^N)$.
The number of each type, $\alpha$, is denoted as $\{N_\alpha\}$ with $\sum_\alpha N_\alpha = N$.
The partition function is
\begin{equation}
  \label{eq:Z}
  Z = \sum_{\{N_\alpha\}} \prod_{\alpha} \frac{1}{N_{\alpha}!}\left(\frac{e^{\beta\mu_{\alpha}}}{\Lambda_{\alpha}^{3}}\right)^{N_\alpha} \iint e^{-\beta \left[U\left(\bm{\xi}^{{N}} \right) + PV\right] } \, d\bm{\xi}^N dV,
\end{equation}
where $\mu_\alpha$ is the chemical potential of each type and $\Lambda_\alpha = \sqrt{\beta h^2 / 2 \pi m_\alpha}$ with mass $m_\alpha$, Planck's constant $h$, and $\beta=1/k_B T$.

As detailed in Ref. \cite{walsh26}, unconstrained MC samples a system with variable $N$ particles by introducing $M-N$ non-interacting fictitious particles.
The extended system, $\bm{\xi}^M$, contains $M$ choose $\{ N_\alpha \}$ copies of each real microstate, $\bm{\xi}^{N\subset M}$, such that integration can be expressed as
\begin{equation}
  \label{eq:dxM}
  \int \pi(\bm{\xi}^M) d\bm{\xi}^M = \sum_{\{ N_\alpha \}}^M \frac{M!}{(M-N)! \prod_\alpha N_\alpha! } \int \pi(\bm{\xi}^{{N} \subset M}) \, d\bm{\xi}^M,
\end{equation}
where $\int d\bm{\xi}^M = \int d\bm{x}^{M-N} d\bm{\xi}^N = V^{M-N} \int d\bm{\xi}^N$ as fictitious particles have only one type.
Sampling the extended system with
\begin{equation}
  \label{eq:piN}
  \pi(\bm{\xi}^{N \subset M}) = \frac{(M-N)!}{M!} \frac{V^N}{V^M} \prod_{\alpha} \left( \frac{e^{\beta \mu_\alpha} }{\Lambda_{\alpha}^{3}}\right)^{N_\alpha} \! \! \! e^{-\beta \left[ U\left(\bm{\xi}^{{N} \subset M}\right) + PV \right]}
\end{equation}
recovers the partition function of Eq. (\ref{eq:Z}) for $N \le M$.
In Metropolis--Hastings MC, the acceptance probability for a move from state $a$ to $b$ is
\begin{equation}
  \label{eq:Aab}
  A_{ab} = \min\left(1,\frac{W_{ba}}{W_{ab}} \frac{ \pi(\bm{\xi}_b^{N_b \subset M})}{\pi(\bm{\xi}_a^{N_a \subset M})} \right),
\end{equation}
where $W_{ab}$ and $W_{ba}$ are the forward and reverse trial probabilities.

Insertion, deletion, and displacement trials were generated using Hamiltonian trajectories as described in Ref. \cite{walsh26}.
Each insertion trial randomly selected one type, which was considered for all fictitious particles.
Specific particle $j$ was selected with biased probability $W^{+j} \propto e^{-\beta^{+} U_j^{+}}$, where $U_j^{+}$ is the hypothetical particle energy considering real neighbors and $\beta^{+}$ is an adjustable parameter.
Deletions began by selecting a random type, which limited the considered atoms.
Specific atom $i$ was chosen with biased probability $W^{-i} \propto e^{\beta^{-} U_i}$ for real atomic energies $U_i$.
Acceptance probabilities are ultimately equivalent to the single-component case with type-dependent $\mu_\alpha$ and $\Lambda_\alpha$.

Configurational sampling was performed with MC trials that selected initial ($\alpha_1$) and final ($\alpha_2$) atomic types with uniform probabilities, $W_{\alpha_1}=W_{\alpha_2}$.
A specific type-$\alpha_1$ atom, $i$, with potential energy $U_i$, was then selected with biased probability
\begin{equation}
  W_{\alpha_1}^{i} = \frac{ e^{\beta^{'}  U_i} }{ \sum_{k}^{N_{\alpha_1}} e^{\beta^{'} U_{k}}},
\end{equation}
where $\beta'$ is an adjustable bias parameter that recovers uniform selection in the limit of $\beta' \rightarrow 0$.
The bias (here, $\beta' = \beta/3$) steered trials toward higher-energy atoms, especially interface sites that required significantly more sampling than the bulk, though the overall effect on acceptance rates was modest.

For selected $\alpha_1$, $\alpha_2$, and $i$, trial states could be generated in two ways.
First, simple direct transmutations impose a Metropolis--Hastings trial probability ratio of $W_{\alpha_2}^i/W_{\alpha_1}^i$, where $W_{\alpha_2}^i$ is the likelihood of a reverse trial.
Alternatively, atom $i$ could be gradually transformed from $\alpha_1$ to $\alpha_2$ over the trajectory of the time-dependent Hamiltonian,
\begin{equation}
  H(t) = [1-\lambda(t)] H_{\alpha_1}^i + \lambda(t) H_{\alpha_2}^i,
\end{equation}
where $H_{\alpha_1}^i$ and $H_{\alpha_2}^i$ respectively describe systems in which particle $i$ is type $\alpha_1$ and $\alpha_2$.
In this case, the trial probability ratio also includes the standard initial velocity term \cite{mehlig92,walsh26}.
Gradual moves were not required to simulate W-V, but significantly help sampling in systems with larger atomic size mismatch.

\subsection{Bulk equilibrium}

W-V was modeled using the Finnis--Sinclair potential of Ref. \cite{chen20}, which is related to the embedded-atom method (EAM).
Individual GB structures are not assumed to be physical---kites appear to be a common weakness of EAM potentials \cite{chen25}---but the general picture is expected to be reasonable.
For instance, grain interiors remained solid solutions across all the considered temperatures and compositions.

Bulk lattice parameters and compositions were calculated using a $\Delta\mu PT$ ensemble that varied $N_\alpha$ for constant $N$ via chemical potential difference $\Delta\mu = \mu_2 - \mu_1$.
2000-atom cells were equilibrated under zero pressure using volume trials that isotropically incremented dimensions on the interval $[-0.05,0.05]$\,\AA\ with probability $W^V=0.005$.
Particle type swaps were attempted with $W'=0.95$.
Displacement trajectories were trialed for 50 5\,fs timesteps with the remaining probability, $W^{0} = 0.045$. 
Simulations were equilibrated until trajectories accumulated $2\cdot10^5$ timesteps and subsequently sampled over $5\cdot10^5$ timesteps.

\subsection{Interface equilibrium}

GBs were modeled as bicrystals with periodic cross-sections.
Boundary area was fixed according to the bulk lattice parameter, $a$, while the nonperiodic normal dimension could vary freely.
Each grain terminated 12$a$ from the interface with the final 6\,\AA\ frozen as a rigid block that contributed a static term to the potential energy.

The $\Sigma27(552)[1\bar{1}0]$ boundary employed a $10\times5$ reconstruction of the $[1,\bar{1},0]$ by $1/2[1,1,\bar{5}]$ cross-section with a total of 8900 atoms.
The $\Sigma5(210)[001]$ and $\Sigma5(310)[001]$ boundaries were modeled similarly with $10\times5$ tilings of the $[0,0,1]$ by $[1,2,0]$ or $[1,3,0]$ periodic units, with 5400 and 7600 atoms respectively.
The $\Sigma41(001)[001]$ boundary used a $2\times2$ reconstruction of the $[5,4,0]$ by $[\bar{4},5,0]$ cell with 7872 atoms.
Real atoms were supplemented with an initially equal number of fictitious particles, i.e. $M=2N$.

Volume trials displaced the upper surface block normal to the interface in the interval $[-0.1,0.1]$\,\AA, while scaling non-rigid positions, with probability $W^{\perp} = 0.0025$.
Alignment trials displaced the block by 0.02\,\AA\ within the interface plane with equal probability, $W^{\parallel} = 0.0025$.
Type swaps were attempted with probability $W'=0.9$ and bias $\beta' = \beta/3$.
Insertions and deletions used trajectories of 200 2\,fs timesteps that were attempted with $W^{+} = W^{-} = 0.025$, bias $\beta^{+} = \beta/5$, and $\beta^{-}=\beta$.
Displacement trajectories of 50 2\,fs timesteps were attempted with the remaining probability, $W^0 = 0.045$.
Most parameters were chosen based on Ref. \cite{walsh26}---a more systematic investigation is planned.

To ensure that the entire range of interface densities was accessible, a quarter of atoms were removed from one half of the initial interface plane, leaving $n=0.75$, while three quarters were removed from the other, leaving $n=0.25$.
Structures were then relaxed so that vacancies would not be immediately filled.
Simulations were then conservatively equilibrated over $5\cdot10^7$ timesteps before sampling over another $5\cdot10^7$ timesteps, which was more than necessary in most, though not all, cases.
GB structures were dethermalized with a short energy minimization before rendering images in OVITO \cite{stukowski10}.
The opacity of bulk atoms was reduced to highlight the boundary structure.
Dislocations were identified per Ref. \cite{stukowski12}.

\section{Data availability statement} The code used in this study is available at \href{https://github.com/flwalsh/lammps-umc}{https://github.com/flwalsh/lammps-umc}.

\begin{acknowledgments}
This work was performed under the auspices of the U.S. Department of Energy by Lawrence Livermore National Laboratory under Contract DE-AC52-07NA27344.
The project was supported by the U.S. Department of Energy, Office of Science under an Office of Fusion Energy Science Early Career Award and partly supported by the LLNL Laboratory Directed Research and Development (LDRD) program under project tracking code 26-LW-125.
Computational resources were provided by the LLNL Institutional Computing Grand Challenge program and the Oak Ridge Leadership Computing Facility at the Oak Ridge National Laboratory, which is supported by the U.S. Department of Energy, Office of Science under Contract No. DE-AC05-00OR22725.
\end{acknowledgments}

\end{document}